\documentclass[11pt,twoside]{article}
\usepackage{asp2010}

\resetcounters

\newcommand{\msun}{M$_\odot$}
\newcommand{\rsun}{R$_\odot$}

\begin{document}

\title{New results on planetary nebula shaping and stellar binarity}
\author{Orsola De Marco
\affil{Macquarie University}
}

\begin{abstract}
The question of what physical mechanisms shape planetary nebulae into their observed morphologies remains open. However, intensified efforts since the last meeting in this series, Asymmetrical Planetary Nebulae IV, in July 2007 have yielded some excellent results. In this review we concentrate on those developments that have taken place in the last three years, with emphasis on results obtained since the review by De~Marco (2009). 
\end{abstract}

\section{The Problem of Shaping a non-Spherical PN}


Approximately 80\% of all planetary nebulae (PN) exhibit morphologies that diverge greatly from a spherical shape \citep{Parker2006}.  Despite advancements in the last 10 years, a convincing answer to what the shaping agent might be is still lacking \citep[for a review see][]{DeMarco2009}. Stellar or even sub-stellar companions have the capacity to interact with upper asymptotic giant branch (AGB) stars and shape the ejected envelope either by strong interactions such as common envelopes \citep{Paczynski1976,Soker1989b} or wider binary interactions such as wind accretion and gravitational focussing \citep[e.g.,][]{Mastrodemos1998,Mastrodemos1999,Edgar2008}. 

However, the fraction of stellar companions to the progenitors of AGB stars that may interact with them is of the order of 30\% \citep{Duquennoy1991}, so how can the fraction of non spherical PN be as high as 80\%? This discrepancy could be explained if  not all the 1-8~\msun\ stars result in a PN. \citet{Moe2006} and \citet{Moe2010} argued, based on population synthesis, that only $\sim$20\% of intermediate mass stars make a PN, with the remainder transiting between the AGB and white dwarf (WD) phases with invisible, or under-luminous nebulae. 

\citet{Soker2005} predicted that deep searches could find the brightest among these under-luminous PN and that they would be spherical (and of course that spherical PN have no binaries in their centres). This prediction has been partly borne out by the MASH survey \citep{Parker2006,Miszalski2008}, the deepest PN survey to date that doubled the fraction of spherical PN from $\sim$10\% to $\sim$20\% and by the {\it Deep Sky Hunters} survey, that found a similar fraction in the very faint population  \citep{Jacoby2010}. These are, in the binary hypothesis, the bright end of the under-luminous, spherical PN formed by single stars and non-interacting binaries. 

A main priority established by the community during the {\it Asymmetrical PN IV} meeting has therefore been to find more binaries in order to relate their parameters to the PN morphology and to determine the PN binary fraction and period distribution. A great deal of success has been enjoyed in the former search (see \S~\ref{sec:list}), while progress has been slow in the latter (\S.

\section{The growing (and shrinking) binary list}
\label{sec:list}

In Table~1 we present new PN that have a highly probable or confirmed binary central stars (for the list of previously known ones see \citealt{DeMarco2009}). Below, we discuss some of them, as well as some that did not make the list, but are likely binaries.

$\bullet$ \citet{Hajduk2010} announced the first binary,  Wolf-Rayet type ([WC7]) central star. Its PN (PN~G221.8-04.2, aka PHR~J0652-0951 and PM~1-23) is similar to A~63, showing an edge-on waist of what may have been a bipolar structure. For assumed primary and secondary masses of 0.6 and 0.3~\msun, respectively, the orbital separation would be $\sim$3 or 5~\rsun, depending on whether the variability is from irradiation (P=0.63~days) or ellipsoidal effects (P=1.2~days). The primary Roche lobe radius would be at 1.3 or 2.2~\rsun. Although the primary radius would be $\sim$0.3~\rsun\ (this is the radius of NGC~40, a slightly hotter [WC8] star \citep{Leuenhagen1996}), a [WC] stellar atmosphere actually extends past the nominal radius value \citep[see, e.g.,][]{DeMarco1998}, so the primary should be filling its Roche lobe.  Of the 33 [WC] and ``week emission line stars" studied by \citet{Hajduk2010}, only this PN revealed a periodically-variable central star, from which the authors concluded that the short-period binary fraction among the hydrogen-deficient central stars may be lower than for the hydrogen-normal ones. This may be in line with a predominant merger origin of these stars \citep{DeMarco2008b} or indicate that the observational biases that affect the [WC] class are different. 

$\bullet$ NGC~6804 and NGC~7139 have a strong IR excess \citep{Bilikova2009} that may be indicative of a companion. We have not included them in Table~1 because a fit to the data imply ``secondary" temperatures of only 1500~K, too cool for a late M companion and possibly too bright for a sub-stellar companion. This temperature, however, is at the condensation limit for dust and possibly unbelievable. Better data are needed to constrain the fits.  

$\bullet$ \citet{Frew2010c} introduced a new class of PN that have a high density, unresolved nebulosity coincidental with the central star. They called this class EGB6-like objects because this PN was the first to have such nebulosity identified. Interestingly, this PN central star was later resolved by the {\it Hubble Space Telescope (HST)} to have a companion and the nebulosity was found to be around the companion, not the central star \citep{Bond2009}. Other objects were recently detected to have such high density unresolved nebulae, usually from the fact that the [OIII] line at 4363~\AA\ is stronger than the H$\gamma$ line at 4340~\AA, indicative of high densities. These include NGC~6804 (De~Marco et al., in preparation), Bran~229 \citep{Frew2010b}, M~2-29 \citep{Gesicki2010}, PHR~J1641-5302 \citep{Parker2003,Frew2010c}, A~57 and PHR~J1553-5738 (Miszalski et al., these proceedings), to name a few. Approximately a dozen EGB6-like objects are known today. Both \citet{Bond2009} and \citet{Gesicki2010} argue that such high density material may be distributed in a disk like structure, or else it would disperse. In the case of EGB6, this disk would be around the companion and would have formed by accretion of AGB primary envelope gas. It is likely that such structures have arisen by binary interactions. 

$\bullet$ NGC~2346 \citep[a binary,][]{Mendez1981}, M~2-29 \citep[a suspected binary,][]{Hajduk2008,Gesicki2010} and CPD-56\,8032 (a [WC10] with a detected dusty disk; \citealt{DeMarco2002b}) have lightcurves showing deep irregular declines, possibly associated with dust emission or a patchy dusty disk. 

$\bullet$ \citet{Wesson2008} discovered a PN around the V~458~Vul (nova Vul 2007), the second such association after GK~Per \citep{Bode1987}. \citet{RodriguezGil2010} determined the period of the binary to be 0.068~days, the shortest period known for a binary central star, and found the binary to be composed of a WD primary and a post-AGB secondary. Another line of evidence associates novae with PN: the high abundance of neon in the hydrogen-deficient ejecta of A30 and A58 \citep{Wesson2003,Wesson2008} interpreted by \citet{Lau2010}  as the products of a nova explosion that took place shortly after the final helium shell flash \citep{Herwig2001}.  

$\bullet$ Santander-Garcia et al. (these proceedings) determined that the secondary star in the binary central star of PN Hen~2-428 is evolved. In agreement with Hillwig (these proceedings), they also determined that about a third of the known central star binary sample \citep{DeMarco2008c,Miszalski2009} has degenerate companions, likely too large compared to the predictions of Moe and De Marco (2010) of $\sim$5\%. 

$\bullet$ Hajduk et al. (these proceedings) announced the detection of a 0.2-mag sinusoidal variation with a period of 20.1 days for the central stars of PN G249.8-02.7 (PHR~J0755-3346), the longest period irradiated central star binary. The variability amplitude appears too large for such a long period, unless the secondary's radius is quite large and/or the central star very hot. We have not listed this PN in Table~1 because the star may not be associated to the nebula (Miszalski, private communication). The binary in the middle of PN K~1-16 (Table~1) also has a long (21.3 days) photometric period, which may or may not be caused by the binary motion.

$\bullet$ \citet{Ostensen2010} detected periodic light variability in the central star of the PN DSH~J1919.5+4445 (Patchick 5; \citealt{Jacoby2010}) a faint, high excitation, elliptical PN with a hint of bipolarity. The asymmetric lightcurve has a period of 1.1 days and an amplitude of only 0.05 magnitude, the smallest known. The central star is a hot subdwarf. Such small amplitude could only be explained with an almost pole-on viewing angle and/or a small secondary radius.

$\bullet$ A35 was demoted from PN status by \citet[][see also Frew\& Parker 2010]{Frew2008b}. It is more likely to be a Str\"omgren sphere around a binary star comprising a G subgiant and a hot component that has left the AGB relatively recently. Finally, four PN from the list of \citet{DeMarco2009} should not be considered binaries until more data is obtained: PHR~J1744-3355, PHR~J1801-2718, PHR~J1804-2645 and PHR~J1804-2913 (Miszalski et al. these proceedings).
\begin{table}
\def~{\hphantom{0}}
  \begin{center}
  \caption{New or updated binary central stars (update on \citealt{DeMarco2009}).}
  \label{tab:knownbinaries}
  \begin{tabular}{lllll}\hline
      PN name     &Type$^1$ &Period&Morph.$^2$ & Reference \\
                         & &(days)   &  &  \\
                         \hline
V~458 Vul            &S1 &0.068&B:W:&\citealt{RodriguezGil2010}\\
Te~11                   & El:&0.12 &E& Miszalski et al. these proceedings\\
NGC~6778         &I:& 0.15 &BPJ& Miszalski et al. these proceedings\\
He~2-428            &El& 0.18 &RW& Santander-Garcia et al., these proceedings\\
K~6-34                  & I& 0.20 &  B:RJ: & revised by \citet{Miszalski2009b}\\
Lo~16                  &EcI& 0.49 &PJ& \citealt{Frew2008b} and Frew et al., in prep.\\
ETHOS~1            & I&0.53 &BJ& Miszalski et al. 2010, submitted\\
PM~1-23 & I: & 0.63 & W & \citealt{Hajduk2010}\\
Necklace$^3$            &I &1.16 &RJB&\citealt{Corradi2010}\\
MPA~J1508-6455 &I: &12.50 &B& Miszalski et al. these proceedings\\
A~14 &Cool,S1: & ? & BR& De~Marco et al., in preparation\\
Bran~229 & Cool & ? &R:P & \citet{Frew2010c}, Frew et al., in prep. \\
A~70 &Cool,UV&?&R&Miszalski et al., in preparation\\
K~1-6                  &Cool,UV & ? &E&\citet{Frew2010}\\
\hline
\multicolumn{5}{l}{$^1$Legend: S1: single-lined spectroscopic binaries: El: ellipsoidal variability; I: irradiated;}\\ 
\multicolumn{5}{l}{Ec: eclipsing; Cool: only a cool stars is known in the system; UV: a hot component is identified}\\
\multicolumn{5}{l}{in the UV. ``:" means that the designation is uncertain.}\\
\multicolumn{5}{l}{$^2$Legend: E: elliptical or indistinct; B: clear, bipolar lobes; R: clear ring(s); W: very likely that PN}\\ 
\multicolumn{5}{l}{is the  edge-on waist of a faded bipolar; J: presence of one or a pair of jets or jet-like structures. }\\
\multicolumn{5}{l}{P: point symmetry. ``:" means that the designation is uncertain.}\\
\multicolumn{5}{l}{$^3$IPHASXJ194359.5+170901.}\\

\end{tabular}
\end{center}
 \end{table}

\section{The PN Binary Fraction and Period Distribution}
\label{sec:binaryfraction}

So far we know that 12-21\% of all PN have post-common envelope central stars, with periods $\la$3~days \citep{Miszalski2009}. This number is not the definitive central star binary fraction for several reasons: (i) The photometric variability technique to detect these binaries is biased to short periods (likely shorter than about 2 weeks; \citealt{DeMarco2008c}). We still do not know how many binary central stars have periods longer than $\sim$2~weeks. (ii) The survey of \citet{Miszalski2009} was affected by a brightness bias and was carried out only in the Galactic Bulge. (iii) Some of the binaries detected by them have been later questioned (Miszalski et al., these proceedings). (iv) Finally, the period distribution of the binaries found by \citet{Miszalski2009} implies that there is a dearth of binaries in the 3 day to 2 week period range. It is however possible that post-CE binaries with periods in the 3-day to 2 week gap are more plentiful than found by \citet{Miszalski2009}, but their irradiation properties may be different due, for instance, to the lack of synchronisation of the orbital and spin period of the secondary for these slightly longer period binaries. This would leave the entire secondary irradiated, reducing the contrast between day and night sides. Although this sounds plausible, the period distribution of central stars is similar to that found via radial velocity technique for the WDs by \citet[][see also Hillwig, these proceedings]{Schreiber2009}. That technique would not suffer this bias.


Radial velocity surveys of central stars of PN are affected by dramatic wind variability that induces spectral line changes that masks even strong periodic binary signals \citep{DeMarco2004,DeMarco2007}.  So the best method to determine the binary fraction with the least number of biases is to test for near-IR excess of a volume-limited sample. With this method we cannot detect periods, although we can get an approximate idea of the companion mass. This method detects unresolved binaries: for a PN at 1 kpc, the orbital separation can be as wide as 500~AU. As a result we will include central star binaries whose separation may be too wide for an interaction having taken place. After detecting the binary fraction in this way, it may not be trivial to account for the fraction of these binaries that has suffered an interaction.

In Fig.~1 (kindly provided by M. Moe) we show the detectability of companions in the $I$ and $J$ bands. From this figure it is clear that to detect faint companions one has to use intrinsically faint central stars. Precision photometry is also needed such that the PN has to be faint in order to afford good background subtraction. The $J$ band is more sensitive, but logistically problematic (IR and optical photometry of the same targets is needed), such that observing in the $I$ band is a more practical approach. Finally, the $H$ band can provide some confirmation as well as an idea of the spectral type of the companion. While the $H$ band can be contaminated by hot dust the $J$ band is unlikely to be. 


We have initiated a search using the NOAO 2.1 m telescope in 2007. So far we have found $I$-band excesses indicative of companions brighter than M5V, in 19-42\% of the central stars of PN observed (5-11 out of 26) at the 3-1$\sigma$ level (Passy et al., in preparation). \citet{Frew2007} used the 2MASS and DENIS databases showing that 53\% of 34 objects in a volume-limited sample have a $J$-band excess down to $\sim$M6V, at the 2$\sigma$ level. Combining these results and de-biasing them to include companions down to the M9V limit using the companion mass distribution for the WD population \citep{Farihi2005}, we conclude that (52$\pm$10)\% of all central stars have a stellar companion closer than $\sim$500~AU. This is tantalisingly higher than predicted by the current PN formation scenario (35\%; \citealt{Duquennoy1991}). However, with a small sample size we cannot call this a solid result, because small number statistics reject the classical scenario with only 1-2$\sigma$ confidence.

\begin{figure}
\vspace{6cm}
\includegraphics{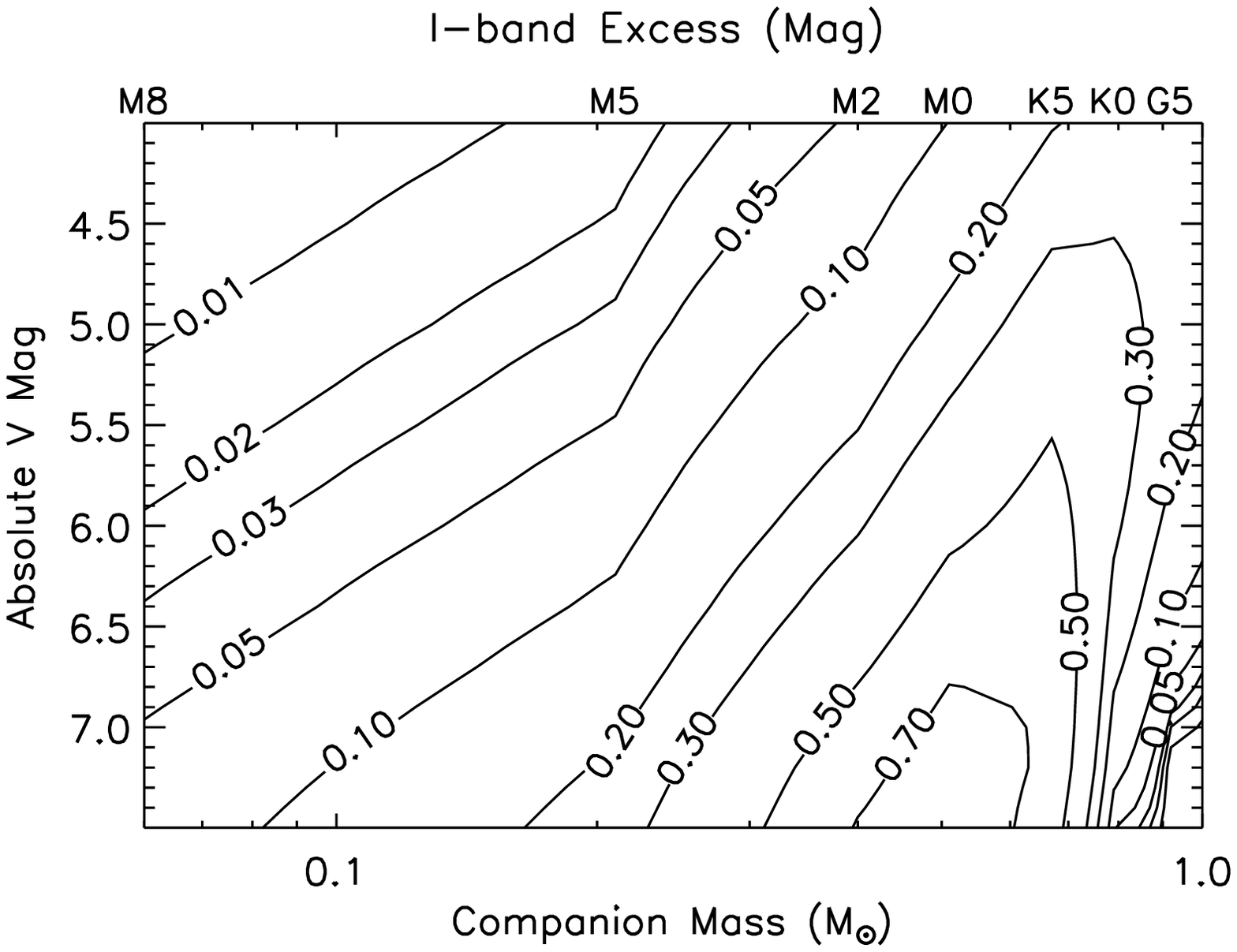}
\includegraphics{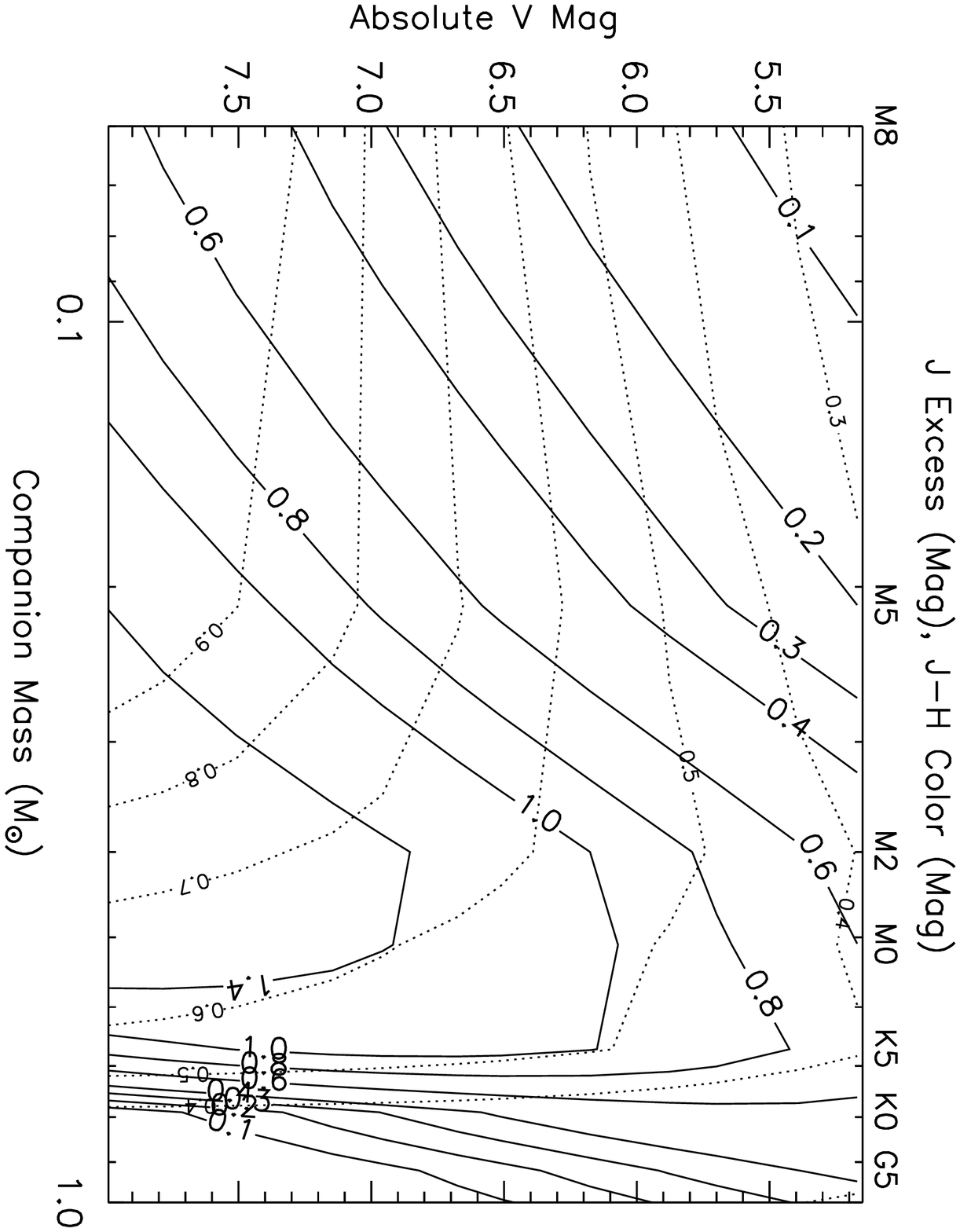}
\caption{Contour plots of the $I$-band (left) and $J$-band (right, solid) excesses expected for companions of different spectral types and for primaries of different intrinsic $V$ brightness ($M_V$). The $J$ band offers increased companion detection sensitivity over the $I$ band. The decrease in $I$/$J$ excess for earlier companions is due to a contamination of the $V$ band by the companion that results in a higher implied reddening resulting in turn in higher predicted (single star) $J$ magnitudes and lower excess. The degeneracy in companion spectral type can however be resolved using the $J - H$ colour (dashed contours). {\it Figure courtesy of M. Moe}}
\end{figure}

The period distribution of central stars of PN will be very hard to determine. For the short period central star binaries known to date a histogram of the known periods can be found in Miszalski et al. (these proceedings) and Hilliwg (these proceedings). Like the post-CE binary period of WDs \citep{Schreiber2009}, the post-CE period for central star of PN peaks at shorter periods than is predicted by any theory of common envelope evolution \citep[e.g.,][]{Davis2010}.
 

\section{Shaping PN with Planets and Brown Dwarfs}

Brown dwarfs and super-Jupiter companions at separations in the range $\sim$2-30~AU may also shape PN \citep[e.g.,][]{Soker1996}. The actual limits are extremely uncertain. Companions closer than the lower limit will interact on the red giant branch and either be unavailable to interact later on during the AGB evolution or even prevent the AGB  ascent altogether. Companions farther than the upper limit have no influence on the AGB star.

We already know that the reservoir of brown dwarfs may be small (also called the brown dwarf desert; \citealt{Grether2006}) although it is slightly larger for larger separations \citep[$\sim$10\%][]{Metchev2008}. However, we do not know how frequently planetary companions at the appropriate orbital separations exist around intermediate mass stars. \citet{Bowler2010} determined that (26+9)\% of stars having a main sequence mass of 1.5 $<$ M/\msun\ $<$ 2.0 host massive planets at large separations (but still less than 3 AU). The new finding is not only of a larger fraction of planet-hosting stars, but also puts the planets at larger orbital separations, where they can more easily be available to shape winds from AGB stars.

\citet{DeMarco2009} pointed out that aside from our ignorance of planetary companion frequency, we also did not know what effect, if any, such low mass companions would have on an AGB star. \citet{Geier2009} discovered an  8-23-M$_{\rm J}$ in a 2.4-day period orbit around the post-red giant branch subdwarf B star HD149382. This companion must have been in a common envelope with its primary and been able to eject the evnelope. 

Based on the morphological considerations of \citet{Soker1997} and \citet{Soker2005} and the population study of \citet{Moe2006}, as well as from the most recent planetary statistics \citet{DeMarco2010} argued that the fraction of PN that have been shaped by planets is of the order of 20\%. If we consider that only 20\% of all intermediate mass stars make PN, then one would also conclude that the fraction of intermediate mass stars that interact with a planet-mass companion on the AGB is of the order of 4\%.

\section{What can be achieved by Asymmetrical Planetary Nebula VI}

At the current rate of discovery, by the next Asymmetrical PN conference in 2013 we should be able to have:

$\bullet$ Precision (1-2\%) photometry of 50-75 central stars of PN in the $B$, $V$ and $I$-band as well as accurate $J$-band photometry for one third to one half of them. This will refine considerably the current, extremely imprecise estimate of the overall PN binary fraction.

$\bullet$ A doubling of the sample of central stars binaries form $\sim$50 to $\sim$100. Most of them will be short period binaries as they are easier to detect and will therefore provide only a biased view of the PN binary population. However such a large sample will allow us to draw statistical conclusions regarding the association between binarity and bipolar morphology (see initial results by \citet{Miszalski2009b})

$\bullet$ At least 20 objects with determined masses and inclinations via radial velocity analysis and stellar atmosphere modelling. There are currently only 6 central stars of PN for which such parameters are known. A larger sample can help characterise the efficiency of the common envelope ejection \citep{DeMarco2010c,Zorotovic2010}.

\acknowledgements
David Frew, Brent Miszalski, Marcin Hadjuk, Todd Hillwig, George Jacoby and Geoff Clayton are thanked for helpful comments. Max Moe is thanked for Figure 1.

\bibliography{../../bibliography}

\end{document}